\documentclass[11pt]{article}
\usepackage[margin=1.25in]{geometry}
\usepackage[usenames,dvipsnames]{color}
\usepackage{url}
\usepackage[colorlinks = true,
            linkcolor = blue,
            urlcolor  = blue,
            citecolor = blue,
            anchorcolor = blue]{hyperref}

\usepackage{graphicx,amsmath,amsfonts,amscd,amsthm,amssymb,pstricks}
\usepackage{color}% Include figure files
\usepackage{dcolumn}% Align table columns on decimal point
\usepackage{bm}% bold math
\usepackage{longtable}
\usepackage{mathrsfs}
\usepackage{textcomp}
\usepackage{esvect}
\usepackage{float}
\usepackage[USenglish,british]{babel}
\usepackage{environ}
\usepackage{xifthen}
\usepackage{xargs}		% programming: better newcommand
\usepackage{hyperref}
\usepackage[separate-uncertainty = true]{siunitx}
\DeclareSIUnit\barn{b}
\usepackage{physics,mathtools}
\usepackage{multirow,tablefootnote}
\usepackage{comment}
\usepackage{enumerate}
\usepackage{appendix}
\usepackage{tikz}	
\usetikzlibrary{arrows, calc, matrix, patterns, decorations.markings, shapes, decorations.pathmorphing, shadows.blur}
\usepackage[utf8]{inputenc}
\usepackage{enumitem}

\def\Title#1{\begin{center} {\LARGE #1 } \end{center}}
\def\Author#1{\begin{center}{ \sc #1} \end{center}}
\def\Address#1{\begin{center}{ \it #1} \end{center}}

\newenvironment{Abstract}{\begin{quotation} \begin{center}
                       ABSTRACT
     \end{center}\bigskip  }{\end{quotation}}

\newcommand\snowmass{\begin{center}\rule[-0.2in]{\hsize}{0.01in}\\\rule{\hsize}{0.01in}\\
\vskip 0.1in Submitted to the  Proceedings of the US Community Study\\ 
on the Future of Particle Physics (Snowmass 2021)\\ 
\rule{\hsize}{0.01in}\\\rule[+0.2in]{\hsize}{0.01in} \end{center}}

\begin{document}

% \pubblock

\Title{Future Circular Lepton Collider FCC-ee: Overview and Status}

\bigskip 

\Author{I.~Agapov$^{1}$, M.~Benedikt$^{2}$, A.~Blondel$^{3}$, M.~Boscolo$^{4}$, O.~Brunner$^{2}$, M.~Chamizo Llatas$^{5}$, T.~Charles$^{6}$, D.~Denisov$^{5}$, W.~Fischer$^{5}$,
E.~Gianfelice-Wendt$^{7}$, 
J.~Gutleber$^{2}$, P.~Janot$^{2}$, M.~Koratzinos$^8$, R.~Losito$^{2}$,
S.~Nagaitsev$^{7,9}$, K.~Oide$^{2,10}$,  T.~Raubenheimer$^{11}$, R.~Rimmer$^{12}$, J.~Seeman$^{11}$, 
D.~Shatilov$^{2}$, V.~Shiltsev$^{7}$, 
M.~Sullivan$^{11}$, U.~Wienands$^{13}$, 
F.~Zimmermann$^{2}$}

\medskip

\Address{$^1$ Deutsches Elektronen-Synchrotron (DESY), Notkestraße 85, 22607 Hamburg, Germany \protect\\ 
$^{2}$ CERN, Esplanade\ des Particules 1, 1211 Geneva 23,
Switzerland \protect\\ 
$^{3}$ U.~Geneva, 24 rue du G\'{e}n\'{e}ral-Dufour, 1211 Geneva 4 Switzerland \protect\\
$^{4}$ INFN-LNF,  
Via Enrico Fermi 54, 00044 Frascati, Roma, Italy \protect\\ 
$^{5}$ Brookhaven National Laboratory, Upton, N.Y., U.S.A. \protect\\
$^6$ University of Liverpool, Liverpool L69 3BX, United Kingdom \protect\\
$^{7}$ Fermi National Accelerator Laboratory, Batavia, Illinois, U.S.A. \protect\\
$^{8}$ Massachusetts Institute of Technology, 
% 77 Massachusetts Ave, 
Cambridge, Massachusetts, U.S.A. \protect\\
$^{9}$ University of Chicago, Chicago, Illinois, U.S.A.\protect\\
$^{10}$ High Energy Accelerator Research Organization, Tsukuba, Ibaraki, Japan \protect\\
$^{11}$ SLAC National Accelerator Laboratory, Stanford, California, U.S.A.\protect\\
$^{12}$ Thomas Jefferson National Accelerator Facility, Newport News, Virginia, U.S.A.\protect\\
$^{13}$ Argonne National Laboratory, Lemont, Illinois, U.S.A.\protect\\
}

\medskip

 \begin{Abstract}
\noindent 
The worldwide High Energy Physics community widely agrees that the next collider should 
be a Higgs factory. 
Acknowledging this priority, in 
2021 CERN has launched the international 
Future Circular Collider (FCC) Feasibility Study (FS).
The FCC Integrated Project foresees, in a first stage, a high-luminosity high-energy electron-positron collider, serving as Higgs, top and electroweak factory, and, in a second stage, an energy frontier hadron collider, with a centre-of-mass energy of at least 100 TeV. 
In this paper, we address a few key elements of the FCC-ee accelerator design, its performance reach, and underlying technologies, as requested by the Snowmass process. 
The Conceptual Design Report for the FCC, published in 2019,   
serves as our primary reference. 
We also summarize a few  recent changes and improvements.
\end{Abstract}

\snowmass

\def\thefootnote{\fnsymbol{footnote}}
\setcounter{footnote}{0}

\section{FCC Integrated Project}
\subsection{Overview}
The Future Circular Collider (FCC) shall 
be located in the Lake Geneva basin and 
linked to the existing CERN facilities
\cite{Benedikt:2781345}.   
The FCC ``integrated programme'' is inspired by the successful past Large Electron Positron collider (LEP) and Large Hadron Collider (LHC)  projects at CERN. It represents a 
comprehensive long-term programme maximising physics opportunities. A similar project is under study in China \cite{thecepcstudygroup2018cepc,ZimmermannPoS}. 
In 2021, CERN has launched the FCC Feasibility Study (FS), that will address not only the technical aspects of the accelerators, 
but also, and in particular, the feasibility of 
tunnel construction and technical infrastructures, 
and the possible financing of the proposed 
future facility. 
The FCC FS is organized as an international collaboration with, presently, about 150 participating institutes from around the world. 
The FCC FS and a future project will profit from 
CERN's decade-long experience with successful large international accelerator projects, e.g., the LHC and HL-LHC, and the associated global  
experiments, such as ATLAS and CMS, to all which
the US has made essential contributions. 
The US participation in CERN based accelerators and experiments during the past decades has been of great mutual benefit.

The first stage of the FCC integrated project is an e$^+$e$^-$ collider, called FCC-ee, which would    
serve as Higgs factory, electroweak and top factory at highest luminosities, and run at four different centre-of-mass energies, namely on the Z pole, at the WW threshold, at the ZH production peak, and at the ${\rm t}\bar{\rm t}$ threshold.
In a second stage, the FCC-ee 
would be followed by a highest-energy proton collider, 
FCC-hh with a centre-of-mass energy of 100 TeV, 
that would naturally succeed the LHC at the energy frontier.
This hadron collider can also accommodate 
ion and lepton-hadron collision options,
providing for  complementary physics. 
The lepton and hadron colliders would
profit from a common civil engineering and 
also from sharing the technical infrastructures.
In particular, 
the FCC would build on and reuse CERN’s existing infrastructure, e.g., the existing chain of 
hadron accelerators, from Linac4 over PSB, PS and SPS to the LHC, can serve as an injector complex for the FCC-hh.

\begin{figure}[htb]
\begin{center}
\includegraphics[width=0.95\linewidth]{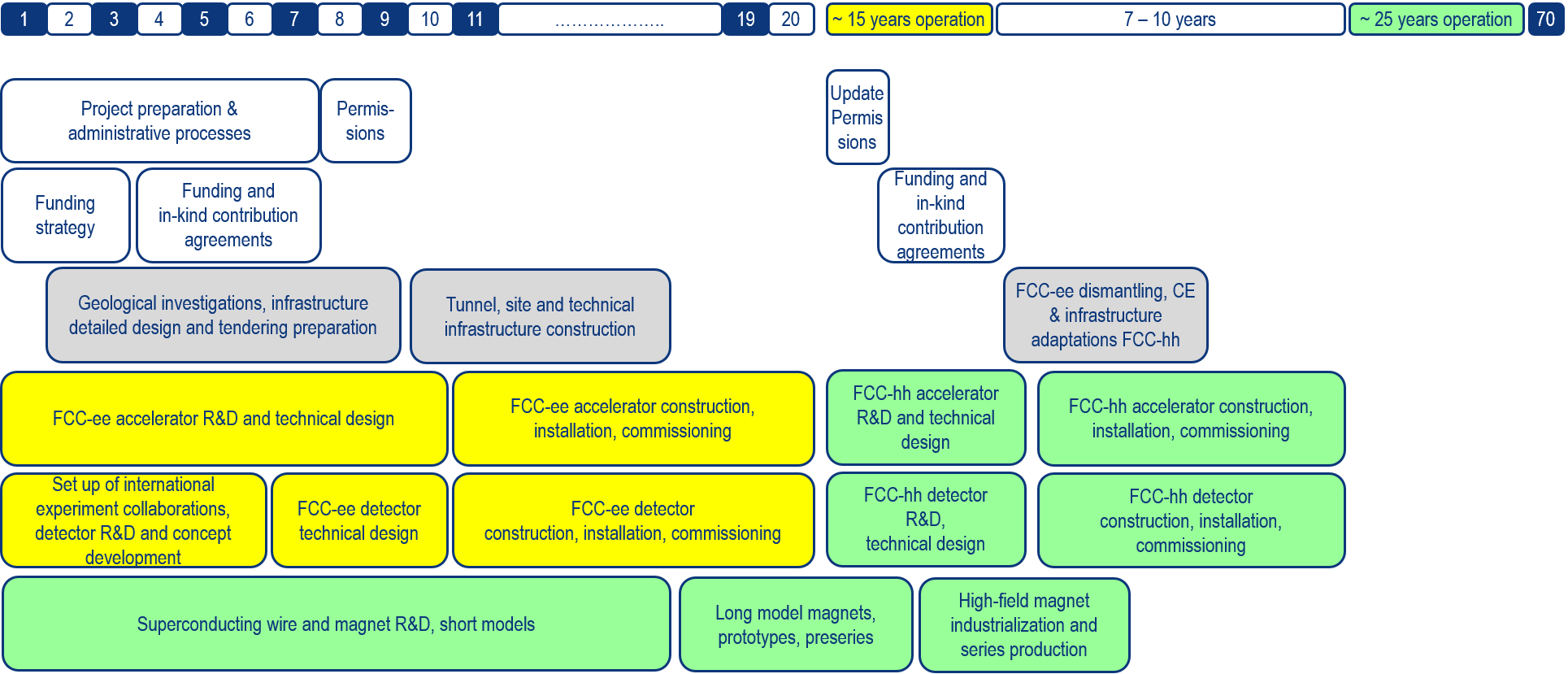}
\vspace*{-3 mm}
\end{center}
\caption{Technical schedule of the FCC integrated project
with year 1 equal to 2021 (a similar schedule was presented 
in Ref.~\protect\cite{naturefcc}).}   
\label{fccschedule}
\end{figure}

The technical schedule of the FCC integrated project 
foresees the start of FCC tunnel construction around
the year 2030 --- or three years 
after a possible project approval ---, the first   e$^+$e$^-$ collisions at the FCC-ee in the early  
2040s, and the first FCC-hh hadron 
collisions by 2065--70 --- see Fig.~\ref{fccschedule}. 
The FCC integrated project would allow 
for a seamless continuation of 
High Energy Physics (HEP) after the
completion of the High Luminosity LHC (HL-LHC) physics programme.

A comprehensive Conceptual Design Report (CDR) for the FCC 
was published in 2019 \cite{fccphys,fccee,fcchh}, describing the 
physics cases, the design of the lepton and hadron colliders,
and the underpinning technologies and infrastructures.

\subsection{Performance} 
The FCC-ee is designed to collide beams at four different  energies, 
with luminosities ranging from
$\sim 2 \times 10^{36}$~cm$^{-2}$s$^{-1}$ 
per Interaction Point (IP), 
or close to $10^{37}$~cm$^{-2}$s$^{-1}$ of 
total luminosity in case of four experiments,   
on the Z pole (91 GeV c.m.),
to about $1.5 \times 10^{34}$~cm$^{-2}$s$^{-1}$ 
per IP 
at the ${\rm t}\bar{\rm t}$ threshold.
Key parameters are summarized in Table 
\ref{tab:perf}.
Thanks to resonant depolarisation, at the two lower
energies, a precision energy calibration is possible, 
down to 100 keV accuracy for $m_Z$ and 300 keV for $m_W$. 

The electrical power
consumption depends on the centre-of-mass energy. 
It, therefore, varies throughout the entire operation
period of the project, extending from about 260 MW to 
 350 MW. 
 These values can be compared with CERN's present 
 power consumption of about 200 MW,  
 when LHC is operating, or with a total 
 CERN power consumption of up to $\sim$240 MW 
 at the time of the previous LEP collider.  
An estimation of the upper limit of the power 
drawn by the various FCC-ee systems \cite{Aull:2156972} 
during regular luminosity production is indicated
in the right-most column of 
Table \ref{tab:perf}.  The numbers shown 
include 37 MW for cooling and ventilation,
36 MW for general services, 8 MW for 
two experiments, 4 MW for data centres, 
and 10 MW for the injector complex.
Although the FCC-ee is three to four times larger than LEP, 
and achieves about $10^{5}$ times the LEP luminosity,
the design concept leads to an overall electrical energy
consumption of only about 2.5 times the one of LEP, which consumed  $\sim$120 MW. 
With additional technology 
advancements and optimisations during operation, 
the overall energy consumption is expected 
to remain close to 300 MW even
at highest collision energies \cite{fccee}.

\begin{table}[htbp]
\caption{Performance figures of FCC-ee.
The ongoing Feasibility  Study allows for a 4 IP collider 
ring with a total integrated luminosity higher by
almost a factor 2.  
For the Z pole (or ${\rm t}\bar{\rm t}$) running the regular
integrated luminosity is show in the table,
while over the first 
two (one) years, the luminosity production  
is expected to be, on average, 
about 2 times lower. 
}
\label{tab:perf}
\begin{center}
\begin{tabular}{lcccc}
\hline\hline
c.m.~energy &
lum./ IP 
& int.~lum./year (2 IPs)
& run time  & power 
\\  { }
[GeV] &
[$10^{34}$ cm$^{-2}$s$^{-1}$]
&  [ab$^{-1}$/yr] 
&  [yr] & [MW]
\\
\hline
91 & 200 & 48 & 4 & 259 \\ 
160 & 20 & 6 & 1--2 & 277  \\ 
240 & 7.5 & 1.7 & 3 & 282 \\ 
365 & 1.3 & 0.34 & 5 & 354 \\ 
\hline\hline    
\end{tabular}
\end{center}
\end{table}

\begin{figure}[bht]
 \includegraphics[width=.9\columnwidth]{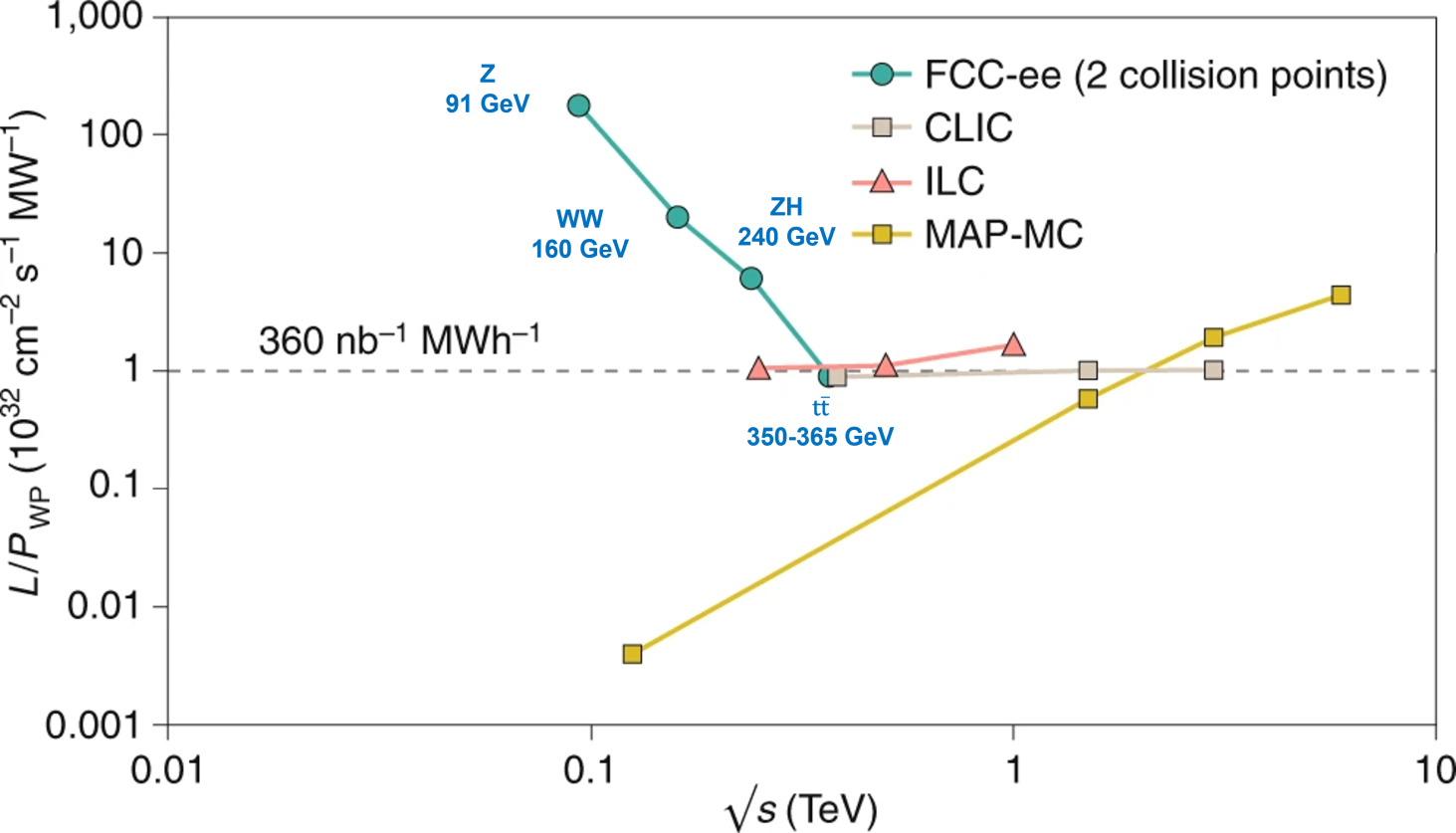}
\caption {\label{fig:eecomp}Luminosity $L$ per supplied electrical wall-plug power $P_{\rm WP}$ is shown as a function of centre-of-mass energy 
for several proposed future lepton colliders 
\protect\cite{Ellis:2691414,naturefcc,Zimmermann:2798333}. 
The FCC-ee electricity cost 
per Higgs boson is about 200 Euro, assuming a price of 
50 Euro MWh$^{-1}$ \protect\cite{fccee,naturefcc}.}
\end{figure}

Based on the CDR parameters, 
the FCC-ee 
is the most sustainable of
all the proposed Higgs and electroweak 
factory proposals, 
in that it implies the lowest energy consumption
for a given value of total integrated luminosity \cite{naturefcc}, 
over the collision energy 
range from 90 to 365 GeV; 
see Fig.~\ref{fig:eecomp}.

\subsection{European Strategy Update 2020 and Feasibility Study Launch} 
The 2020 Update of the European Strategy for Particle Physics (ESPPU) \cite{esppu} 
states that ``\textit{An electron-positron Higgs factory is the highest-priority next collider. For the longer term, the European particle physics community has the ambition to operate a proton-proton collider at the highest achievable energy.}" and 
“\textit{Europe, together with its international partners, should investigate the technical and financial feasibility of a future hadron collider at CERN with a centre-of-mass energy of at least 100 TeV and with an electron-positron Higgs and electroweak factory as a possible first stage. 
Such a feasibility study of the colliders and related infrastructure should be established as a global endeavour and be completed on the timescale of the next Strategy update.}”
Responding to this key request from the ESPPU, 
in the summer of 2021, the five-year Future Circular Collider 
Feasibility Study was launched \cite{council-fcc-1,council-fcc-2}. 

\subsection{Collider Design Optimization} 
\subsubsection{Placement and revised layout}
\begin{figure}[htb]
\begin{center}
\includegraphics[width=0.75\linewidth]{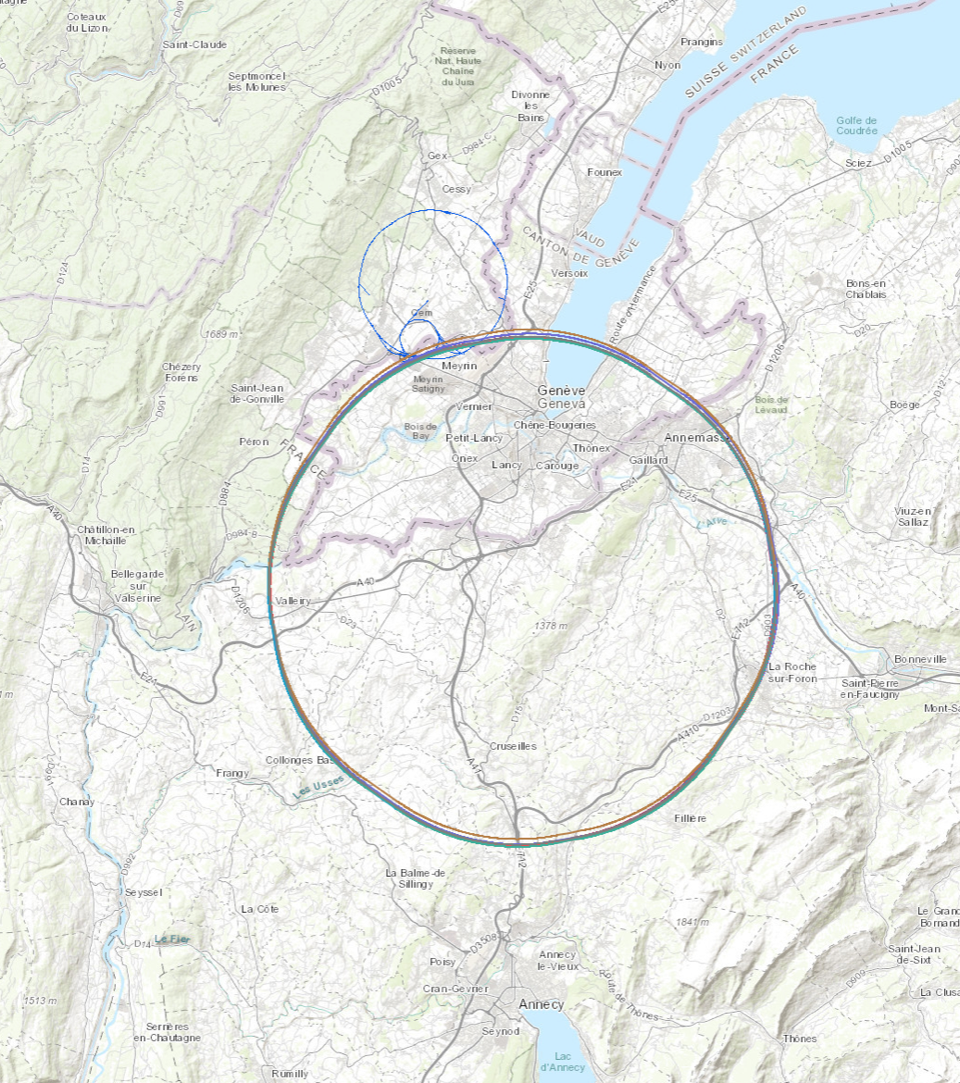}
\vspace*{-3 mm}
\end{center}
\caption{Some of the FCC implementation variants presently under study. The existing 
SPS and LHC rings are shown for  reference.}  
\label{FCCscenarios}
\end{figure}

The 2019 FCC CDR describes the baseline FCC  
design with a circumference 
of 97.75 km, 12 surface sites,
and two primary collision points.
In 2021, a further design optimisation has 
resulted in an optimised placement of much  
lower risk, with a circumference of about 
91.2 km and only 8 surface sites,  and 
which would be compatible with
either 2 or 4 collision points. 
A few of the optimised FCC implementation variants 
currently under study are depicted 
in Fig.~\ref{FCCscenarios}.  

\begin{figure}[htb]
\begin{center}
\includegraphics[width=0.65\linewidth]{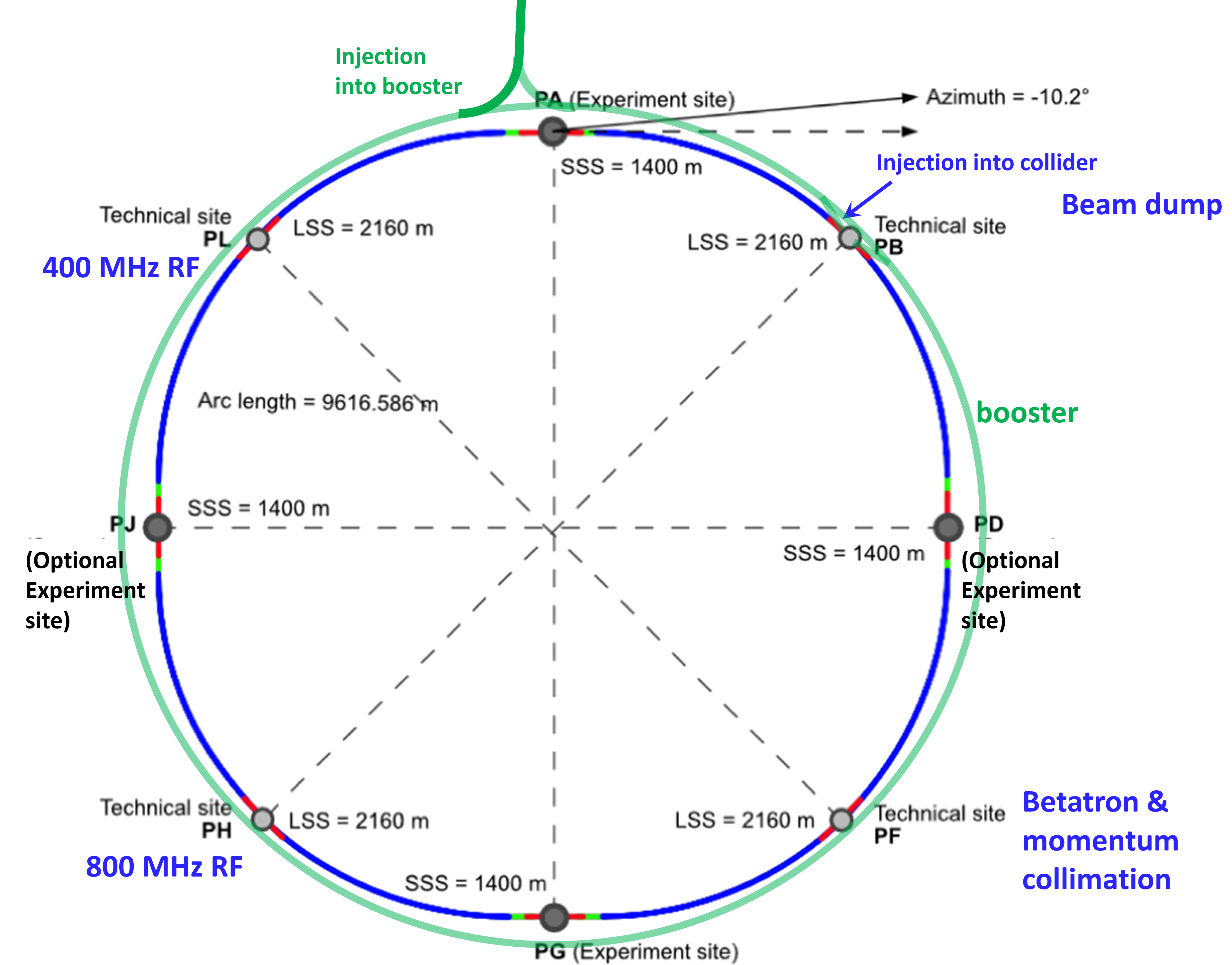}
\vspace*{-3 mm}
\end{center}
\caption{Schematic layout of FCC-ee and its booster with a circumference of 91.2 km and strict four-fold superperiodicity. }  
\label{fcclayout}
\end{figure}

Consequently, adaptations of the CDR design
and re-optimisation of the machine parameters
are underway, taking into account
not only the new placement, but also, for FCC-ee,  
the possibly larger number of interaction points, 
and the mitigation of complex
``combined'' effects, e.g.~the interplay
of transverse and longitudinal impedance
with the beam-beam interaction. 
Figure ~\ref{fcclayout} sketches 
the layout 
and possible straight-section functions for the 
electron-positron collider.
A consistent layout for the hadron collider
can be found in the accompanying 
White Paper for FCC-hh.

\subsubsection{Parameter update}
A preliminary table with key FCC-ee parameters for the cases of 
either two or four IPs is shown in Table \ref{eefuturetable}. 
In the CDR \cite{fccee}, the operation at the Z and W assumed a 60$^{\circ}$/60$^{\circ}$ phase advance per arc cell. The mitigation of the combined impedance and beam-beam effects requires a larger momentum compaction factor than in the CDR \cite{shatilovpc}. 
This has resulted in a ``long'' 90$^{\circ}$ cell, of twice the cell length used for the H and ${\rm t}\bar{\rm t}$ operation \cite{oidepc}.

\begin{table}[htbp]
\caption{Preliminary key parameters of FCC-ee (K. Oide, 2021),
as evolved from the CDR parameters, now  
with a shorter circumference of 91.2 km,
and a new arc optics for Z and W running. 
Luminosity values are given per interaction point (IP),
for scenarios with either 2 (left) or 4 IPs (right).  
Both the natural rms bunch lengths (b.~lengths) 
and rms relative beam energy spreads (en.~spreads) due to synchrotron radiation (SR) and their 
collision values including 
beamstrahlung (BS) are shown.
The FCC-ee considers a combination of 400 MHz radiofrequency systems  
(at the first three energies, up to $2\times$2 GV) and 800 MHz (additional cavities for ${\rm t}\bar{\rm t}$ operation),
with respective voltage strengths as indicated.
The beam lifetime shown represents the combined effect of
the luminosity-related 
radiative Bhabha scattering and 
beamstrahlung, the latter relevant
only for ZH and ${\rm t}\bar{\rm t}$ running
(beam energies of 120 and 182.5 GeV).
\label{eefuturetable} 
}
\begin{center}
\begin{tabular}{|l|cccc|cccc|}
\hline\hline
Running mode & Z & W & ZH & ${\rm t}{\bar{\rm t}}$ &
  Z & W & ZH & ${\rm t}{\bar{\rm t}}$  
\\ \hline
Number of IPs & \multicolumn{4}{c|}{2} &
 \multicolumn{4}{c|}{4}  
\\ 
%\cline{2-10} 
Beam energy (GeV)& 
45.6 &  80 & 120 &  182.5 &
45.6 & 80 & 120 & 182.5 
\\ 
Bunches/beam & 
12000 & 880 & 272 & 40 &
10000 & 880 & 248 & 36 
\\
Bunch population [$10^{11}$] & 
2.02 & 2.91  & 1.86  & 2.37  &
2.43 & 2.91 & 2.04 & 2.64 
\\
Beam current [mA] &
1280 & 135 &  26.7 & 5.0 & 
1280  & 135 &  26.7 &  5.0
\\
Lum. / IP [10$^{34}$cm$^{-2}$s$^{-1}$]&
193 & 22.0 & 7.73 & 1.31 &
182 & 19.4 & 7.26 & 1.33 \\
Energy loss / turn [GeV] &
0.039  & 0.37 & 1.87  &  10.0 & 
0.039  & 0.37 & 1.87  & 10.0  
\\
Synchr.~Rad.~Power [MW] &
\multicolumn{4}{|c|}{100} & 
\multicolumn{4}{|c|}{100} \\ 
RF Volt.~400 MHz [GV] & 
0.12 & 1.0 & 2.08 & 4.0 & 
0.12 & 1.0 & 2.08 & 4.0 \\
RF Volt.~800 MHz [GV] & 
0 & 0 & 0 & 7.25 & 
0 & 0 & 0 & 7.25 \\
Rms b.~length (SR) [mm]&
	4.38 & 3.55 & 3.34 & 2.02 & 
	4.38 & 3.55  & 3.34 & 2.02  
	\\
	\hspace*{1 cm} (+BS) [mm]&
	12.1 & 7.06 & 5.12 & 2.56 & 
	14.5 &  8.01 &  6.00 & 2.95 
	\\
	Rms en.~spread (SR) [\%]&
	0.039 & 0.069 & 0.103 & 0.157 & 
	0.039 & 0.069 & 0.103 & 0.157 
	\\
	\hspace*{1 cm} (+BS) [\%]&
	0.108 & 0.137 & 0.158 & 0.198 & 
	0.130 & 0.154 & 0.185 & 0.229 
	\\
Rms hor.~emit.~$\varepsilon_{x}$ [nm] & 
0.71  & 2.17 & 0.64 & 1.49 & 
0.71  & 2.17 & 0.64 & 1.49 \\
Rms vert.~emit.~$\varepsilon_{y}$ [pm] & 
1.42 & 4.32 &  1.29 &  2.98 & 
1.42 & 4.32 &  1.29 & 2.98 \\
Norm.~hor.~em.~$\gamma \varepsilon_{x}$ [$\mu$m] & 
63  & 340 & 150 & 530 & 
63  & 340 & 150 & 530  
 \\
Norm.~vert.~em.~$\gamma \varepsilon_{y}$ [$\mu$m] & 
0.13 & 0.68 &  0.30 &  1.06 & 
0.13 & 0.68 &  0.30 &  1.06 \\
Longit.~damp.~time [turns] &
1170 & 216 & 64.5 & 18.5 & 
1170 & 216 & 64.5 & 18.5 \\
Hor.~IP beta $\beta_x^{\ast}$ [mm] & 
100 & 200 & 300 & 1000 & 
100 & 200 & 300 & 1000 \\
Vert.~IP beta $\beta_y^{\ast}$ [mm] & 
0.8 & 1.0 & 1.0 & 1.6 & 
0.8 & 1.0 & 1.0 & 1.6 \\
Beam lifetime [min.] &
35 & 32 & 9 & 16 & 
19 & 18 & 6 & 9 \\
\hline\hline
\end{tabular}
\end{center}
\end{table}

The beam parameters, in particular the emittances, bunch length, lifetime, and luminosity still need to be validated in strong-strong beam-beam simulations and in weak-strong simulations including errors and optics corrections.
The luminosity values per IP are slightly higher for two IPs than for four, but rather similar.   
Therefore, the beam lifetime due to radiative Bhabha scattering (inversely proportional to the total luminosity) is about a factor two higher, which allows a more aggressive choice for the beamstrahlung-induced lifetime.  

\subsubsection{Monochromatization option}
In addition to the 4 baseline running modes
on the Z pole, at the WW threshold, at the (Z)H production peak, and above the ${\rm t\overline t}$ threshold, listed 
listed in Table \ref{eefuturetable}, another optional 
operation mode, presently under investigation 
for FCC-ee, is the direct $s$-channel Higgs production, ${\rm e}^{+}{\rm e}^{-}\rightarrow {\rm H}$,
at a centre-of-mass energy of 125 GeV,
which would allow a direct measurement of the electron Yukawa coupling.  
Here, a monochromatization scheme 
should reduce the effective collision energy spread in order for the latter to become 
comparable to the width of the Higgs \cite{afausepj}.  
Various possible techniques for monochromatization --- e.g., based on nonzero IP
dispersion, residual chromaticity, transversely deflecting cavities or a combination thereof ---  
presently are under investigation. 
The monochromatization option, with 
the necessary development, control and monitoring of the monochromatization level, 
requires novel techniques 
in its own right~\cite{Eliana:2021zix}.

\subsection{Design Challenges}
The FCC-ee builds on 60 years of operating colliding beam storage rings.  The design is robust and will provide high luminosity over the desired centre-of-mass energy range from 90 to 365 GeV. 

Valuable lessons were learnt from the highest energy e$^+$e$^-$ collider
so far, LEP, and from its hadronic  
successor, the LHC, as well as from the two B factories, PEP-II and KEKB.    
Importantly, the SuperKEKB collider, presently being commissioned,  
features many of the key elements of FCC-ee:   
double ring, large crossing angle, low vertical IP beta function $\beta_{y}^{\ast}$ (design value $\sim$0.3~mm),  short design beam lifetime of a few minutes, 
top-up injection, and a positron  production rate of up to several $ 10^{12}$/s.  
SuperKEKB has already achieved, in both rings, the world's smallest ever 
$\beta_{y}^{\ast}$ of $0.8$~mm,
which also is the lowest value 
considered for FCC-ee. 
Profiting from a new 
``virtual'' crab-waist collision scheme,
first developed for FCC-ee \cite{koide},
in June 2021 SuperKEKB reached a world record
luminosity of  $3.81 \times 10^{34}$~cm$^{-2}$s$^{-1}$.

For FCC-ee, two key operational ingredients are 
 ``top-up'' injection in routine operation,
and ``bootstrapping'' injection when restarting, e.g., after a failure  \cite{dmitry}.
Alternating re-injection to top up the circulating electron and positron bunches maintains approximately constant beam current and luminosity, so that the average luminosity
of FCC-ee 
approaches the peak luminosity.
This type of operation mode was pioneered at the PEP-II, KEKB and SuperKEKB colliders.
For the FCC-ee, 
the top-up injection needs to ensure that 
intensities of colliding bunches are kept equal to within a few percent, 
to avoid a beam-beam flip-flop effect,
where one bunch would blow up and the opposite bunch shrink, resulting in an unrecoverable situation.   
For the same reason, when filling 
the machine from zero, a novel
bootstrapping injection process 
is foreseen, with  alternating injections into either of the two collider rings, avoiding large charge imbalances
\cite{dmitry}. 

At the FCC-ee, highly precise energy calibration using resonant depolarisation
at c.m.~energies of 91 and 160 GeV, with wiggler-polarised pilot bunches, 
and roughly $10^{5}$ times higher luminosity than LEP, 
will allow measuring  the masses of the Z and W bosons, as well as the width of 
the Z, with 90 times improved 
precision, rendering the FCC-ee 
an exceptional electroweak factory \cite{blondelprec}.

\section{Technology Requirements}
The main technological systems required for the FCC-ee are the 400 and 800 MHz SRF systems, the energy efficient arc magnets, the arc vacuum system, 
a few special magnets for the interaction region, and the positron source.

\subsection{Technology Readiness Assessment} 
The technology to build a machine like FCC-ee long exists.
Already in 1976 (almost half a century ago !), 
B.~Richter stated that ``An e$^+$e$^-$ 
storage ring in the range of a few hundred 
GeV in the centre of mass can be built with present 
technology'' \cite{Richter:312037}. 

The 400/800 MHz RF systems proposed for FCC-ee are state of the art, 
or close to it.
The 400 MHz Nb/Cu cavities are based on the technology developed  
for the LEP and LHC cavities \cite{VogelSRF99}. 
A first prototype 5-cell 800 MHz bulk-Nb cavity constructed by JLAB 
 met the design specifications \cite{Marhauser:2653853}.  
R\&D is necessary to reduce cost and increase reliability of the cavity production process and operation, 
not to reach the performance. 
For the klystrons the situation is similar.

Resembling modern light sources, the arc vacuum system is based on 
a round copper vacuum chamber featuring winglets 
for photon stops, and thin NEG coating. 
Prototypes of the energy-efficient 
twin-aperture arc dipoles and quadrupoles \cite{PhysRevAccelBeams.19.112401} 
have been built and measured \cite{Milanese:2744311}. 
A prototype of the SC
canted-cosine-$\theta$ final focusing quadrupole has also been fabricated  \cite{Koratzinos:2313367}.  
The positron production rate required from the positron source is comparable to 
those achieved at the SLC and at SuperKEKB.

\subsection{Ongoing and Planned R\&D} 
The FCC-ee technology R\&D is focused on incremental improvements aimed mainly at further optimising efficiency, obtaining the required diagnostic precision, and on achieving the target performance in terms of beam current and luminosity. 
FCC-ee will strive to include new technologies if they can increase efficiency, decrease costs or reduce the environmental impact of the project.    
Key FCC-ee R\&D items for improved energy efficiency  
include high-efficiency continuous wave (CW)
radiofrequency (RF) power sources (klystrons and/or solid state), high-$Q$ 
superconducting (SC) cavities for the 
400--800 MHz range, and possible applications of HTS magnets. 

Aside from the various RF systems, another major
component of the FCC-ee is the regular arc, covering almost \SI{80}{km}.  
The arc cells must be cost effective, 
reliable and easily maintainable. Therefore,  
as part of the FCC R\&D programme it is 
planned to build a complete arc half-cell mock up including girder, vacuum system with antechamber and pumps, dipole, quadrupole and sextupole magnets, beam-position monitors, cooling and alignment systems, and technical infrastructure interfaces, by the year 2025.
An intriguing parallel R\&D path concerns the possibility to realize 
the arc quadrupoles plus sextupoles using HTS.  
Another mock up is proposed for the interaction region, 
consisting, e.g., of the central beam pipe, 
final SC quadrupole, support structures, stabilization system, 
and remotely controlled flanges.

It is also envisaged to design, construct and then test  
with beam a novel positron source 
(including a superconducting solenoid as adiabatic matching device)  
plus capture linac, and measure the achievable positron yield,  
at the PSI SwissFEL facility, with a primary electron 
energy that can be varied from 0.4 to 6 GeV. 

In addition, 
to prepare for ultra-high precision centre-of-mass energy measurements, a
focused R\&D effort is covering  
state-of-art and beyond in terms of spin-polarisation simulations and measurements (inverse Compton scattering, beamstrahlung, etc.). 

Finally, for high luminosity, high current operation of FCC-ee,
advanced beam stabilization/feedback 
systems to suppress instabilities arising over a few turns 
will be developed, based on the systems developed for the B factories, but possibly 
augmented by a narrow band system to combat the low-frequency 
resistive-wall instability.

% A robust low-impedance collimation scheme 
% can be extrapolated from similar systems at 
% SuperKEKB and LHC/HL-LHC.

% Profiting from ongoing smart-automation 
% efforts for the existing CERN accelerators, 
% a machine tuning system based on artificial intelligence (AI)
% is foreseen. 

\subsection{FCC-ee Demonstrators} 
With SuperKEKB an actual demonstrator machine is available.  
As stated, SuperKEKB  features many of the same key ingredients, and its design
parameters are even more challenging than those of FCC-ee.  
The FCC collaboration is working together with the SuperKEKB team to 
address, fully understand and mitigate remaining issues limiting
SuperKEKB performance. 

Beam studies relevant to FCC-ee --- for example on optics correction, vertical emittance tuning, crab-waist collisions, or beam energy calibration --- can, and will, also be conducted at INFN-LNF/\-DAFNE, DESY/PETRA III,  and KIT/KARA \cite{jacqui}.

The US Electron-Ion Collider (EIC) is set to start beam operation in 2030.  
The electron beam parameters of the EIC 
are similar to, and even more demanding than, those of FCC-ee.
For example, the EIC foresees operation with
two times higher beam current than required for the FCC-ee Z pole running.  
Therefore, about a decade before FCC-ee commissioning, the EIC can serve as an important test bed for a variety of FCC-ee components and concepts --- collimators, radiofrequency systems, beam feedbacks, diagnostics, 
interaction region, 
operation with polarised beams, etc.

\section{Staging Options and Upgrades}
It is planned to operate the FCC-ee first on the Z pole (91 GeV c.m., 4 years), then on the W threshold (160 GeV, 2 years), on the ZH production peak (240 GeV c.m., 3 years), and, after a 1 year shutdown, at the ${\rm t}\bar{\rm t}$ threshold (365 GeV, 5 years). In general as the energy is increased and the beam current decreases, additional RF systems are installed with higher RF voltage and higher impedance.  
As mentioned, an additional optional running 
mode at 125 GeV c.m. (direct Higgs production), with monochromatization, for a couple of years.

\subsection{Energy Upgrades}
There appears to be 
no motivation from particle physics to increase the FCC-ee energy beyond
365 GeV c.m. For example, Ref.~\cite{Blondel:2018aan} concludes that
``500 GeV is not a particularly useful energy for the lepton colliders under consideration, especially for the FCC-ee.''
Regardless, in principle, should there emerge a motivation to do so, 
the c.m.~energy of FCC-ee could be raised to 400 GeV or beyond, by pushing the RF voltage or installing additional RF cavities.

\subsubsection{100 TeV hadron collider}
The main upgrade of the FCC-ee consists in its complete disassembly and replacement by a 100 TeV hadron collider, FCC-hh \cite{fcchh}, 
when the high-field magnets needed for the latter are available in series production.
The hadron collider would use the same tunnel infrastructure,
experimental caverns, general services, cooling \& ventilation systems,
and surface sites as the lepton collider. 
It might also reuse the FCC-ee cryoplants, with upgrades and additions.

\subsubsection{ERL option}
An upgrade of e$^+$e$^-$ collisions to higher energies, $\sim 600$~GeV or beyond, has been proposed
through converting the FCC-ee into a few-pass ERL \cite{fcceeerl}. 
The general feasibility of such an upgrade, its compatibility with the FCC-ee layout, and the realistically attainable luminosity require further studies and R\&D. This option is not part of the FCC 
Feasibility Study.

\subsubsection{Muon collider option}
Following the end of the FCC-ee physics programme, its booster ring could also be used to accelerate positrons to 45 GeV, at a high  rate, in order to produce low-emittance muon beams through positron annihilation \cite{Antonelli:2015nla}. Accelerating these muons in the existing CERN complex, and colliding them in the LHC might offer a path towards a 14 TeV muon collider \cite{Zimmermann:2018wfu,Zimmermann-ipac22}.
Achieving the needed luminosity appears quite challenging, but may not be fully excluded by introducing and combining several new concepts.

\subsection{Luminosity Upgrades}
The FCC-ee luminosity already is remarkably high. 
Nevertheless it could be further increased in a number of ways.  
First, with collisions at 4 instead of 2 experiments the total luminosity can be 
increased by about a factor of 1.7.   This option is already part of the FCC Feasibility Study. 
Second, enlarging the dynamic momentum acceptance through optics design improvements or more sophisticated correction schemes would allow for higher luminosity at the same beam lifetime. 
Third, a more powerful injector complex could sustain shorter beam lifetimes, and would, thereby, also support higher luminosity.   
Fourth, increasing the RF power would permit higher beam current, and again higher luminosity, but this is not a preferred path. 
Finally, on paper, also the ERL option promises a high luminosity at the Higgs and ${\rm t}\bar{\rm t}$ energies.

\subsection{Experimental System Upgrades}
Experimental system upgrades for the up to four different detectors 
are presently not foreseen during the 
15 years of planned physics operation.

\section{Synergies with other concepts and/or existing facilities}
Substantial synergies exist with the presently operating LHC 
at CERN and SuperKEKB in Japan, 
with the Electron-Ion Collider (EIC) 
in the United States, 
with the CEPC design in China 
--- which is extremely similar to the FCC-ee ---, 
and with the most modern synchrotron light sources such as 
ESRF-EBS in France, and  
next-generation low-emittance storage rings,  including the APS Upgrade 
in the US, 
and PETRA IV in Germany. 

\subsection{Synergies on Machine Technologies}
The FCC-ee design, optics and operation features great synergies
with other past, present and future circular colliers, including LHC, SuperKEKB, and EIC, 
and with modern low-emittance synchrotron light sources.
The SRF systems and RF power sources 
show additional synergies with other facilities utilizing superconducting RF, including the European XFEL, CEBAF, LCLS-II, PIP-II, etc.
The injector complex equally has strong synergies with the SuperKEKB injector, 
but also with other warm linacs, such as the SwissFEL, FERMI, and PAL-XFEL.

\subsection{Synergies on Detector Technologies}
Various types of detectors are considered for the up to four collision points, which could be 
optimized for different types of physics priorities,
such as ``Higgs Factory'' programme, ``ultraprecise electroweak'' programme, ``heavy flavour'' physics, and the search for ``feebly coupled particles'', respectively
\cite{mdam}.  
There are significant synergies with the developments for the existing LHC experiments and their upgrades, and with developments for the ILC and CLIC detectors. 

\subsection{Synergies on conventional facilities and green power}
Numerous synergies exists and are being further developed. 
Noteworthy are the low-loss transport of electrical power
over long distances, e.g. using HTS cables, 
energy efficient RF power sources, robust superconducting  RF systems, energy efficient magnet systems, including HTS-based magnets, 
development of more efficient cryoplants,  
remote interventions based on robots,
recovery of the spent energy, in terms of heat or electricity, 
and intelligent reuses of the excavated material 
(``mining for the future'' initiative).

\subsection{Synergies for Physics Research}
An enormous 
synergy and complementarity exist between the FCC-ee and FCC-hh physics programmes,
as detailed in Ref.~\cite{fccphys}.
Should both the ILC and the FCC-ee be constructed,  
the operation modes of these two e$^+$e$^-$ 
machines could be adjusted 
and tailored so as to maximise the overall physics output 
\cite{Blondel:2706695}.

\section{Conclusions}
The circular electron-positron collider FCC-ee is a compelling,  
energy-efficient option for the much wanted 
Higgs and electroweak factory.
It may also serve as a key step towards the next energy frontier 
machine, in the form of the subsequent FCC-hh hadron collider. 
The FCC Feasibility Study and a possible 
future FCC project offer 
a unique opportunity for the US to participate,
similar to the LHC, 
but as an even more ambitious and global endeavour, 
to jointly develop advanced technologies with 
important spin offs for society, and 
to tackle several outstanding key  
questions about our universe.

\bibliographystyle{JHEP}
\bibliography{mybib}  % file myreferences.bib

\providecommand{\href}[2]{#2}\begingroup\raggedright\begin{thebibliography}{10}

\bibitem{Benedikt:2781345}
M.~Benedikt and F.~Zimmermann, \emph{{The physics and technology of the Future
  Circular Collider}},
  \href{https://doi.org/10.1038/s42254-019-0048-0}{\emph{Nature Reviews
  Physics} {\bfseries 1} (2019) 238}.

\bibitem{thecepcstudygroup2018cepc}
{The CEPC Study Group}, \emph{{CEPC Conceptual Design Report: Volume 1 -
  Accelerator}}, {\emph{arXiv} {\bfseries 1809.00285} (2018) }.

\bibitem{ZimmermannPoS}
F.~Zimmermann, \emph{{Towards a Circular Higgs and Electroweak Factory}},
  {\emph{{Proceedings of Science, The European Physical Society Conference on
  High Energy Physics (EPS-HEP2021), 26--30 July 2021, Online conference,
  jointly organized by Universit\"{a}t Hamburg and the research center DESY;
  Copyright owned by the author(s) under the Creative Commons License CC
  BY-NC-ND 4.0.}} (2022) }.

\bibitem{naturefcc}
M.~Benedikt, A.~Blondel, P.~Janot et~al., \emph{{Future Circular Colliders
  succeeding the LHC}},
  \href{https://doi.org/{https://doi.org/10.1038/s41567-020-0856-2}}{\emph{Nature
  Physics} {\bfseries 16} (2020) 402}.

\bibitem{fccphys}
{The FCC Collaboration}, \emph{{FCC Physics Opportunities: Future Circular
  Collider Conceptual Design Report Volume 1}}, {\emph{{European Physical
  Journal C}} {\bfseries 79} (2019) }.

\bibitem{fccee}
{The FCC Collaboration}, \emph{{FCC-ee: The Lepton Collider}},
  {\emph{{Eur.~Phys.~J.~Spec.~Top.}} {\bfseries 228} (2019) }.

\bibitem{fcchh}
{The FCC Collaboration}, \emph{{FCC-hh: The Hadron Collider}},
  {\emph{{Eur.~Phys.~J.~Spec.~Top.}} {\bfseries 228} (2019) }.

\bibitem{Aull:2156972}
S.~Aull, M.~Benedikt, D.~Bozzini, O.~Brunner, J.-P.~Burnet, A.~Butterworth
  et~al., \emph{{Electrical Power Budget for FCC-ee}}, .

\bibitem{Ellis:2691414}
R.K.~Ellis et~al., \emph{{Physics Briefing Book: Input for the European
  Strategy for Particle Physics Update 2020}}, {\emph{arXiv} {\bfseries
  1910.11775} (2019) CERN}.

\bibitem{Zimmermann:2798333}
F.~Zimmermann, M.~Benedikt and A.-S.~Müller, \emph{{The Future Circular
  Collider Study}},
  \href{https://doi.org/10.18429/JACoW-IPAC2020-MOVIR01}{\emph{JACoW IPAC}
  {\bfseries 2020} (2020) MOVIR01. 5 p}.

\bibitem{esppu}
{European Strategy Group}, \emph{{2020 Update of the European Strategy for
  Particle Physics (Brochure)}}, {\emph{CERN-ESU-015} (2020) }.

\bibitem{council-fcc-1}
{CERN Council}, \emph{{Organisational structure of the FCC feasibility study.
  Restricted CERN Council - Two-Hundred-and-Third Session}},
  {\emph{CERN/SPC/1155/Rev.2} (2021) }.

\bibitem{council-fcc-2}
{CERN Council}, \emph{{Main deliverables and timeline of the FCC feasibility
  study. Restricted CERN Council - Two-Hundred-and-Third Session}},
  {\emph{CERN/SPC/1161} (2021) }.

\bibitem{shatilovpc}
D.~Shatilov, \emph{{Private communication}}, .

\bibitem{oidepc}
K.~Oide, \emph{{Private communication}}, .

\bibitem{afausepj}
A.~Faus-Golfe, M.~Valdivia~Garcia and F.~Zimmermann, \emph{{The Challenge of
  Monochromatization Direct s-Channel Higgs Production: ${\mathrm e}^+{\mathrm
  e}^- \rightarrow {\mathrm H}$}}, {\emph{{Eur. Phys. J. Plus}} {\bfseries 137}
  (2022) }.

\bibitem{Eliana:2021zix}
A.~Blondel and E.~Gianfelice, \emph{{The challenges of beam polarization and
  keV-scale centre-of-mass energy calibration at the FCC-ee}},
  \href{https://doi.org/10.1140/epjp/s13360-021-02038-y}{\emph{Eur. Phys. J.
  Plus} {\bfseries 136} (2021) 1103}.

\bibitem{koide}
K.~Oide et~al., \emph{Design of beam optics for the future circular collider
  e$^{+}$e$^{-}$ collider rings},
  \href{https://doi.org/10.1103/PhysRevAccelBeams.19.111005}{\emph{Phys. Rev.
  Accel. Beams} {\bfseries 19} (2016) 111005}.

\bibitem{dmitry}
D.~Shatilov, \emph{{FCC-ee Parameter Optimization}}, {\emph{{ICFA Beam Dyn.
  Newsl.}} (2017) 30}.

\bibitem{blondelprec}
A.~Blondel, P.~Janot, J.~Wenninger et~al., \emph{{Polarization and
  Centre-of-Mass Energy Calibration at FCC-ee}}, {\emph{arXiv 1909.12245}
  (2019) }.

\bibitem{Richter:312037}
B.~Richter, \emph{{Very high-energy electron-positron colliding beams for the
  study of the weak interactions}},
  \href{https://doi.org/10.1016/0029-554X(76)90396-7}{\emph{Nucl. Instrum.
  Methods} {\bfseries 136} (1976) 47}.

\bibitem{VogelSRF99}
H.~Vogel et~al., \emph{{Production of Nb/Cu Sputtered Superconducting Cavities
  for LHC}},  in \emph{Proc. 9th Int. Conf. RF Superconductivity (SRF'99)},
  pp.~437--439, JACoW Publishing, Nov., 1999,
  \href{https://jacow.org/SRF99/papers/WEP016.pdf}{https://jacow.org/SRF99/papers/WEP016.pdf}.

\bibitem{Marhauser:2653853}
F.~Marhauser et~al., \emph{{802 MHz ERL Cavity Design and Development}},  in
  \emph{{Proc. 9th Int. Particle Accelerator Conf. (IPAC'18)}}, pp.~3990--3992,
  JACoW Publishing, Apr.-May, 2018,
  \href{{http://accelconf.web.cern.ch/ipac2018/papers/THPAL146.pdf}}{{http://accelconf.web.cern.ch/ipac2018/papers/THPAL146.pdf}}.

\bibitem{PhysRevAccelBeams.19.112401}
A.~Milanese, \emph{Efficient twin aperture magnets for the future circular
  e$^{+}$/e$^{-}$ collider},
  \href{https://doi.org/10.1103/PhysRevAccelBeams.19.112401}{\emph{Phys. Rev.
  Accel. Beams} {\bfseries 19} (2016) 112401}.

\bibitem{Milanese:2744311}
A.~Milanese, J.~Bauche and C.~Petrone, \emph{{Magnetic Measurements of the
  First Short Models of Twin Aperture Magnets for FCC-ee}},
  \href{https://doi.org/10.1109/TASC.2020.2976949}{\emph{IEEE Trans. Appl.
  Supercond.} {\bfseries 30} (2020) 4003905. 5 p}.

\bibitem{Koratzinos:2313367}
M.~Koratzinos, \emph{{The FCC-ee Interaction Region Magnet Design}},
  \href{https://doi.org/10.18429/JACoW-eeFACT2016-TUT1AH3}{\emph{{Proc.~ICFA
  workshop eeFACT2016, Daresbury}} (2017) TUT1AH3. 4 p}.

\bibitem{jacqui}
J.~Keintzel et~al., \emph{{Experimental beam tests for FCC-ee}}, {\emph{{PoS,
  Proc.~EPS-HEP2021}} (2021) }.

\bibitem{Blondel:2018aan}
A.~Blondel and P.~Janot, \emph{{Future strategies for the discovery and the
  precise measurement of the Higgs self coupling}}, {\emph{arXiv} {\bfseries
  1809.10041} (2018) }.

\bibitem{fcceeerl}
V.N.~Litvinenko, T.~Roser and M.~Chamizo-Llatas, \emph{{High-energy
  high-luminosity e+e- collider using energy-recovery linacs}},
  \href{https://doi.org/https://doi.org/10.1016/j.physletb.2020.135394}{\emph{Physics
  Letters B} {\bfseries 804} (2020) 135394}.

\bibitem{Antonelli:2015nla}
M.~Antonelli, M.~Boscolo, R.~Di~Nardo and P.~Raimondi, \emph{{Novel proposal
  for a low emittance muon beam using positron beam on target}},
  \href{https://doi.org/10.1016/j.nima.2015.10.097}{\emph{Nucl. Instrum. Meth.
  A} {\bfseries 807} (2016) 101}
  [\href{https://arxiv.org/abs/1509.04454}{{\ttfamily 1509.04454}}].

\bibitem{Zimmermann:2018wfu}
F.~Zimmermann, \emph{{LHC}/{FCC}-based muon colliders},
  \href{https://doi.org/10.1088/1742-6596/1067/2/022017}{\emph{Journal of
  Physics: Conference Series} {\bfseries 1067} (2018) 022017}.

\bibitem{Zimmermann-ipac22}
F.~Zimmermann, A.~Latina, A.~Blondel, M.~Antonelli and M.~Boscolo, \emph{{Muon
  Collider based on Gamma factory, FCC-ee and Plasma Target}},
  {\emph{{Submitted to IPAC'22}} (2022) }.

\bibitem{mdam}
M.~Dam, \emph{{Detector R\&D requirements for future circular high energy e+e-
  machines}}, {\emph{{Input session of Future Facilities I, ECFA R\&D Roadmap
  Input, February 2021}} (2021) }.

\bibitem{Blondel:2706695}
A.~Blondel and P.~Janot, \emph{{Circular and Linear e$^+$e$^-$ Colliders:
  Another Story of Complementarity}}, {\emph{arXiv} {\bfseries 1912.11871}
  (2019) }.

\end{thebibliography}\endgroup

\end{document}